\documentclass{article}
\usepackage{spconf, amsmath, amssymb, graphicx}
\usepackage{xcolor}
\usepackage{hyperref}
\usepackage{bm}

\usepackage{mathtools}
\usepackage{booktabs}

\usepackage{subcaption}
\usepackage{graphics}
\usepackage{caption}
\usepackage{wrapfig}

\usepackage[export]{adjustbox}
\usepackage{algorithm}
\usepackage{algorithmic}
\usepackage{multirow}
\usepackage{float}

\title{Prompting and Adapter Tuning for Self-Supervised Encoder-Decoder Speech Model}

\name{%
\begin{tabular}{@{}c@{}}
Kai-Wei Chang$^{1}$ \qquad 
Ming-Hsin Chen$^{2}$ \qquad 
Yun-Ping Lin$^{2}$ \qquad 
Jing Neng Hsu$^{2}$ \qquad \\
Paul Kuo-Ming Huang$^{2}$ \qquad
Chien-yu Huang$^{1}$ \qquad 
Shang-Wen Li$^{3}$ \qquad
Hung-yi Lee$^{1}$
\end{tabular}
}
\address{$^1$Graduate Institute of Communication Engineering, National Taiwan University, Taiwan, \\$^2$ Department of Computer Science and Information Engineering, National Taiwan University, $^3$Meta AI}

\copyrightnotice{979-8-3503-0689-7/23/\$31.00~\copyright2023 IEEE}
\begin{document}
%
\maketitle
\begin{abstract}
Prompting and adapter tuning have emerged as efficient alternatives to fine-tuning (FT) methods. However, existing studies on speech prompting focused on classification tasks and failed on more complex sequence generation tasks. Besides, adapter tuning is primarily applied with a focus on encoder-only self-supervised models. Our experiments show that prompting on Wav2Seq, a self-supervised encoder-decoder model, surpasses previous works in sequence generation tasks. It achieves a remarkable 53\% relative improvement in word error rate for ASR and a 27\% in F1 score for slot filling. Additionally, prompting competes with the FT method in the low-resource scenario. Moreover, we show the transferability of prompting and adapter tuning on Wav2Seq in cross-lingual ASR. When limited trainable parameters are involved, prompting and adapter tuning consistently outperform conventional FT across 7 languages. Notably, in the low-resource scenario, prompting consistently outperforms adapter tuning.

\end{abstract}
\begin{figure*}[t]
    \centering
    \includegraphics[width=.9\linewidth]{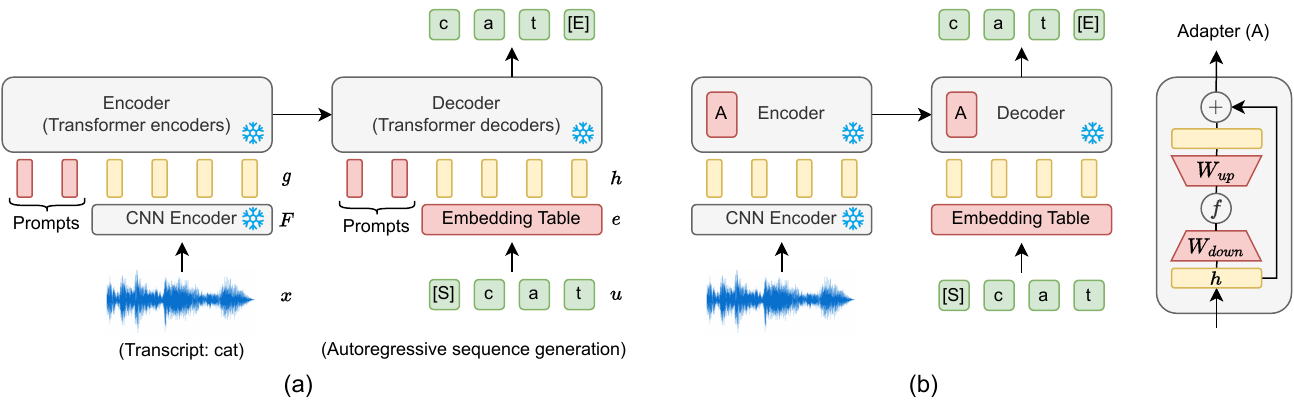}
    \caption{The proposed prompting and adapter tuning on encoder-decoder SSL speech models.
    Red components are trained during downstream task adaptation.
    (a) Prompting: Trainable prompts are prepended at the input of Transformer layers in both encoder and decoder parts.
    (b) Adapter tuning: Adapters are inserted in the Transformer layers.}
    \label{fig:wav2seq}
\end{figure*}

\section{Introduction}
\label{sec:intro}
Self-supervised learning (SSL) enables models to learn informative representations from unlabeled data and has achieved impressive results in various domains and tasks.
Notable examples are CPC~\cite{oord2018representation}, Wav2vec2.0~\cite{baevski2020wav2vec}, HuBERT~\cite{hsu2021hubert} and so on.
These pre-trained SSL speech models often act as feature encoders and a downstream model is built on top of them to perform downstream tasks~\cite{DBLP:conf/interspeech/YangCCLLLLSCLHT21, mohamed2022self}.
On the other hand, several works \cite{lakhotia2021generative, kharitonov2022text, hassid2023textually} proposed using offline clustering methods (e.g., K-means) on these speech representations to obtain discrete speech tokens.
Decoder-only models are then trained on top of these speech tokens to perform generative tasks.
Wav2Seq~\cite{wu2023wav2seq} further combines these methods, in which the encoder and decoder are both pre-trained.
In the pre-training stage, Wav2Seq is trained to perform ASR on a pseudo language, and this has been shown to benefit downstream sequence generation tasks.
However, despite the architectural and pre-training variations, these models usually require fine-tuning to adapt to specific downstream tasks effectively.
Due to a large number of parameters, the conventional \textbf{fine-tuning (FT)} approach on these models becomes costly regarding computational resources.

To this end, \textbf{prompting}~\cite{liu2023pre} and \textbf{adapter tuning}~\cite{houlsby2019parameter} have gained popularity as effective approaches in addressing the aforementioned challenges.
Prompting involves utilizing prompts, which are templates or task-specific vectors, to guide the pre-trained model. By fitting the input data into these prompts, the model is steered towards better understanding or generation, enabling it to perform various downstream tasks effectively~\cite{liu2023pre, chang2022exp}.
Different from making modifications at the input side, adapter tuning inserts lightweight neural networks into a pre-trained model. 
These components, referred to as \textbf{adapters}, are then trained while fixing the remaining parameters of the pre-trained model.
Adapter tuning has shown to be effective in mitigating the domain mismatch problem in transfer learning~\cite{houlsby2019parameter}.
Overall, both techniques offer the advantage of significantly reducing the number of task-dependent parameters compared to conventional FT approaches, thus alleviating computational resource limitations.

While prompting and adapter tuning techniques have been explored in the field of speech processing, previous research has only primarily focused on either decoder-only or encoder-only pre-trained models.
SpeechPrompt~\cite{chang2022exp, chang2023speechprompt} applies prompting to a decoder-only GSLM~\cite{lakhotia2021generative} on various speech classification tasks but fails in sequence generation tasks.
Other works~\cite{otake2023parameter, chen2023exploring} have focused on adapter tuning in encoder-only speech representation learning models, such as HuBERT~\cite{hsu2021hubert}, wav2vec2.0~\cite{baevski2020wav2vec}, and WavLM~\cite{chen2022wavlm}. 
On the other hand, recent studies~\cite{wu2023wav2seq, ao2022pre} have demonstrated that encoder-decoder pre-trained models are more suitable for sequence generation tasks, such as ASR and speech translation.
However, it has not yet been explored whether these models can efficiently transfer to other downstream tasks through prompting and adapter tuning.
This paper presents the first application of prompting and adapter tuning techniques for the encoder-decoder pre-trained model.
As Fig.~\ref{fig:wav2seq} shows, we perform prompting and adapter-tuning on Wav2Seq, an SSL encoder-decoder model, and achieve superior performance to SpeechPrompt in terms of both classification and sequence generation tasks.
This paper is also the first to achieve non-trivial results in prompting speech models for sequence generation tasks.
More specifically, it achieves 53\% relative improvement in WER for ASR and 27\% in F1 score for slot filling.
We also conducted extensive studies comparing prompting, adapter tuning, and fine-tuning methods in cross-lingual transfer learning across 7 different languages.
Both prompting and adapter tuning achieve superior parameter efficiency compared to the fine-tuning method.
Moreover, the experiment reveals that in the low-resource scenario, prompting consistently outperforms adapter tuning.

\begin{figure*}[t]
    \centering
    \includegraphics[width=.9\linewidth]{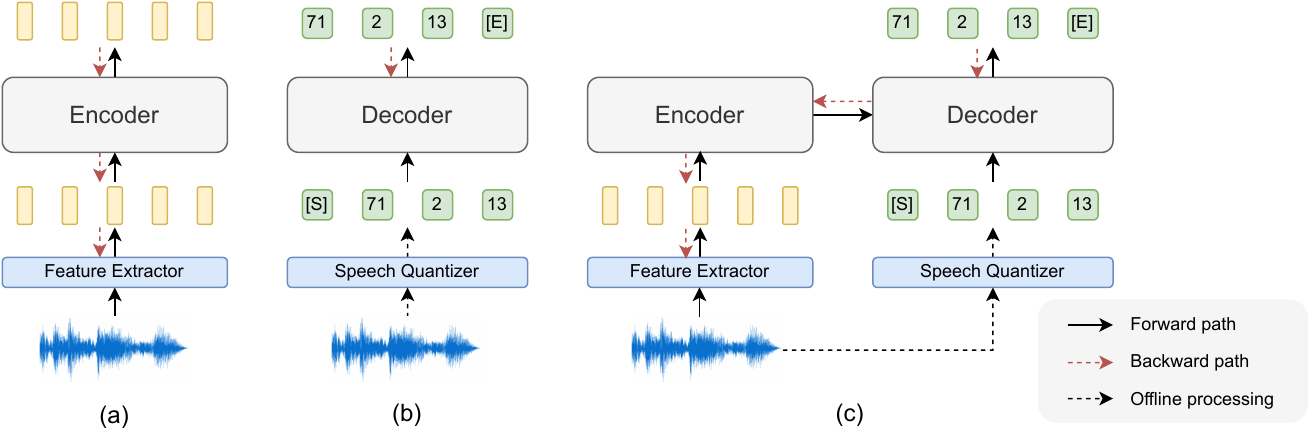}
    \caption{Typical self-supervised speech model architectures. \textbf{(a) Encoder-only models}: They can extract informative representations for downstream speech processing tasks. \textbf{(b) Decoder-only models}: By performing quantization on speech representations, the speech is transformed into discrete speech tokens. A decoder-only model is then trained on top of these speech tokens. Examples include GSLM and TWIST. \textbf{(c) Encoder-decoder models}: Combining the techniques in (a) and (b). The encoder and the decoder are jointly trained. Examples include Wav2seq and Speech2C.}
    \label{fig:speech_models}
\end{figure*}

\section{Related Works}
\label{sec:related-works}
\subsection{Self-supervised speech models}
\label{sec:related-works-ssl-models}
SSL models have achieved state-of-the-art performances across various tasks.
These models can be mainly grouped into three distinct types: encoder-only, decoder-only, and encoder-decoder architectures.
The overview of typical SSL speech models is shown in Fig.~\ref{fig:speech_models}.
Encoder-only models usually serve as representation learning models that can extract features to benefit downstream tasks.
Examples include Wav2vec2~\cite{baevski2020wav2vec}, HuBERT~\cite{hsu2021hubert}, and WavLM~\cite{chen2022wavlm}.
While these models encode speech into informative representations, they cannot be directly used for downstream tasks.
To adapt an encoder-only model to downstream tasks, we need to additionally build a downstream model that utilizes the encoded representations and fine-tunes the whole model~\cite{mohamed2022self}.
On the other hand, decoder-only models are often built for spoken generative language modeling, such as GSLM~\cite{lakhotia2021generative}, AudioLM~\cite{borsos2023audiolm}, and TWIST~\cite{hassid2023textually}.
These models mostly consist of an offline speech quantizer and a generative decoder-only model.
The speech quantizer quantizes the input speech into a sequence of discrete tokens, and the decoder module is then trained with these tokens to perform generative language modeling.
Last, in encoder-decoder models, both encoder and decoder are jointly trained in the pre-training stage~\cite{wu2023wav2seq, ao2022pre, DBLP:conf/interspeech/ArunkumarU22}, which is different from the previous two types.
Encoder-decoder models are superior across various sequence generation tasks, such as ASR and speech translation~\cite{wu2023wav2seq, ao2022pre} and thus we focus on this specific type in this paper.
Wav2Seq~\cite{wu2023wav2seq} and Speech2C~\cite{ao2022pre} share similar model architecture. We adopt the former for its pre-training shows benefits to various tasks, while the latter focuses on ASR.

\subsection{Prompting in speech processing}
\label{sec:related-works-prompting}
The prompting method~\cite{liu2023pre} fits the data into a task-specific template that steers the pre-trained model to perform a given task without modifying its architecture and parameters.
Prompts can take the form of either natural language templates~\cite{petroni2019language, shin2020autoprompt} or trainable prompts~\cite{li2021prefix, liu2022p} positioned at the input of the model.
Trainable prompts can also be placed at the input of each model's layer, referred to as \textbf{deep prompt tuning}~\cite{li2021prefix} to offer more capability to the prompts.
In this paper, we focus on deep prompt tuning as it is efficient and effective to prompt pre-trained speech models~\cite{chang2022exp, wu2023speechgen}.
Another similar approach is model reprogramming~\cite{DBLP:conf/iclr/ElsayedGS19, DBLP:journals/corr/abs-2202-10629}, where a transformation is applied to the input data so that the pre-trained model can be used for a target task.
Recently, model reprogramming methods are also applied to several speech and audio-related tasks, such as spoken command classification~\cite{yen2021neural}, cross-lingual speech recognition~\cite{yang2023english}, and music genre classification~\cite{hung2023low}.

\subsection{Adapter tuning in speech processing}
\label{sec:related-works-pet}
Adapter tuning~\cite{rebuffi2017learning, houlsby2019parameter} is an alternative method of fine-tuning. It utilizes bottleneck neural networks known as adapters to transfer the hidden representation of a pre-trained model to a specific downstream. 
For SSL speech models, there have been explorations of adapters like Houlsby adapter~\cite{houlsby2019parameter}, LoRA~\cite{DBLP:conf/iclr/HuSWALWWC22}, and AdapterBias~\cite{fu2022adapterbias} on encoder-only SSL speech models, as reported in a recent study~\cite{chen2023exploring}.
Moreover, adapter tuning is not confined to SSL models; it is also applicable to conventional speech models for domain adaptation, such as language adaptation in speech translation~\cite{le2021lightweight} and speaker adaptation in text-to-speech~\cite{morioka2022residual}.

\vspace{-0.3cm}
\section{Method}
\label{sec:method}
\subsection{Self-supervised encoder-decoder speech model}
Fig.~\ref{fig:wav2seq} shows a widely used self-supervised encoder-decoder speech model built with Transformer~\cite{vaswani2017attention} blocks.
This paper is also based on this model.
The CNN encoder $F$, serving as a feature extractor, encodes input speech $\bm{x}$ into initial features $\bm{g}^{(1)} = F(\bm{x}) =[g_1^{(1)}, g_2^{(1)}, ..., g_T^{(1)}]$, where $T$ is the sequence length.
The Transformer encoder $E$ consists of multiple Transformer layers and takes $\bm{g}^{(1)}$ to generate contextualized representations.
Specifically, the $i$-th Transformer layer receives $\bm{g}^{(i)} = [g_1^{(i)}, g_2^{(i)}, ..., g_T^{(i)}]$ and output $\bm{g}^{(i + 1)}$.

Similar to the encoder, each layer of the decoder receives $\bm{h}^{(i)} = [h_1^{(i)}, h_2^{(i)}, ..., h_{T'}^{(i)}]$ as input, where $T'$ denotes the sequence length.
In particular, the first decoder representation, $\bm{h}^{(1)}$, involves the utilization of the token embedding table. 
Each token embedding is obtained by transforming a corresponding discrete token using a lookup embedding table  $\bm{e}$. Consequently, $h^{(1)}$ can be written as $\bm{h}^{(1)} = [\bm{e}(u_1), \bm{e}(u_2), ..., \bm{e}(u_{T'})]$. Here, $u_1, u_2, ..., u_{T'}$ denote the tokens present in the decoder sequence.

\subsection{Prompting}
Deep prompt tuning prepends trainable prompt vectors $\bm{p}$ at the input of each Transformer layer.
For example, in the encoder, the input of the $i$-th layer is modified as follows:
\begin{equation}
    g^{(i)} \leftarrow Concat(\bm{p}^{(i)}, g^{(i)})
\end{equation}
where $\bm{p}^{(i)} = [p_1^{(i)}, p_2^{(i)}, ..., p_l^{(i)}]$ is a sequence of prompt vectors of length $l$.
Moreover, in addition to input modification, we also prepend prompt vectors to the key and value in the attention mechanism:

\begin{align}
    K^{(i)} &= Concat(\bm{p}_K^{(i)}, g^{(i)})W_K^{(i)} \\
    V^{(i)} &= Concat(\bm{p}_V^{(i)}, g^{(i)})W_V^{(i)}.
\end{align}
This enhances the model's ability to attend to relevant information and tailor its attention patterns according to the prompts.
The decoder $D$ undergoes similar modifications.
Notably, in prompt tuning, the pre-trained encoder and decoder's architecture and parameters always remain fixed.

\subsection{Adapter tuning}
\label{sec:method-adapter}
Adapter tuning involves the integration of lightweight neural networks into Transformer layers, providing a flexible and efficient way to enhance model performance.
Typically, an adapter consists of a down projection network $\bm{W}_{down} \in \mathbb{R}^{d \times r}$, an up projection network $\bm{W}_{up} \in \mathbb{R}^{r \times d}$, an activation function $f$, and a residual connection.
This design forms a bottleneck structure, wherein the bottleneck dimension $r$ serves as a hyperparameter that can be adjusted to create adapters of varying sizes. The embedding of Transformer layer $\bm{h}$ can therefore be adjusted by the adapter:
\begin{equation}
    \bm{h} \leftarrow \bm{h} + f(\bm{h}\bm{W}_{down})\bm{W}_{up}
\end{equation}
By incorporating adapters into Transformer layers, the model gains additional capacity for task-specific modifications without significantly increasing the overall parameter count.
The down projection network reduces the dimensionality of the input, enabling efficient adaptation to the target task, while the up projection network restores the dimensionality to match the original input size.
The residual connection facilitates the flow of information, ensuring that the original information is preserved while incorporating task-specific adjustments.

\section{Experimental Settings}
We used Wav2Seq~\cite{wu2023wav2seq} as the backbone model and performed prompting and adapter tuning experiments on it.
Firstly, we assessed the effectiveness of prompting on Wav2Seq by comparing its performance with SpeechPrompt~\cite{chang2022exp} in speech processing across several classification tasks and sequence generation tasks.
Furthermore, we applied prompting to Wav2Seq in cross-lingual speech recognition tasks to investigate the transferability of this technique. 
Additionally, we compared the effectiveness and parameter efficiency of adapter tuning and prompting in both low-resource and full-dataset scenarios.
Table~\ref{table:tasks} presents an overview of the tasks and datasets involved in the experiments.

\begin{table}[t]
    \scriptsize\centering
    \caption{The summary of downstream tasks in our experiment.
    \textbf{SC}: speech classification. \textbf{SG}: sequence generation. } 
    \vskip -0.02in
    \label{table:tasks}
    \scriptsize\centering
    \resizebox{\linewidth}{!}{
        \begin{tabular}{lcl}
        \toprule
        \textbf{Task} & \textbf{Type} & \textbf{Dataset}\\
        \midrule
            Keyword Spotting        & SC & Speech Commands~\cite{DBLP:journals/corr/abs-1804-03209}\\
            Intent Classification  & SC & Fluent Commands~\cite{DBLP:conf/interspeech/LugoschRITB19}\\
            ASR - 100 hours  & SG & LibriSpeech-100 hr~\cite{DBLP:conf/icassp/PanayotovCPK15} \\
            ASR - 10 hours & SG & LibriLight-10 hr~\cite{librilight} \\
            Slot Filling   & SG & Audio SNIPS~\cite{DBLP:conf/icassp/LaiCL0G21} \\
            Multilingual-ASR & SG & Multilingual LibriSpeech~\cite{DBLP:conf/interspeech/PratapXSSC20}\\
        \bottomrule
        \end{tabular}
    }
\end{table}
\label{sec:exp}
\subsection{Wav2Seq}
\label{sec:exp_model}
Wav2Seq~\cite{wu2023wav2seq} consists of Transformer encoders and decoders, which are jointly trained with the pseudo speech recognition task.
It first quantizes SSL speech representations into discrete tokens using an offline K-means clustering algorithm.
Repetitive tokens are then merged through a process called deduplication.
Subsequently, a subword tokenization algorithm, namely byte-pair encoding (BPE) is applied to generate the desired pseudo subwords.
These pseudo subwords serve as targets for pseudo speech recognition in the self-supervised pre-training stage, where the encoder and decoder are jointly optimized.
When utilizing Wav2Seq for a downstream task, the embedding table for the model's vocabulary is substituted and fine-tuned specifically for that task.

\subsection{Prompting paradigm}
\label{sec:exp_prompting}
We compared our prompting method to SpeechPrompt~\cite{chang2022exp} in both speech classification tasks and sequence generation tasks.
We set the prompt length to 5 for the classification tasks and 120 for the sequence generation tasks, respectively.
The trainable parameters used align with those in SpeechPrompt.
For classification tasks, we conducted Keyword Spotting (KS) and Intent Classification (IC).
In keyword spotting, models aim to detect the keyword in an utterance by classifying the utterance into a pre-defined keyword set.
In intent classification, models aim to extract and summarize the ``action'', ``object'', and ``location'' from an utterance, which collectively represent the user's intent.
For sequence generation tasks, we conducted ASR and Slot Filling (SF).
While we used LibriSpeech-100 for ASR, we also considered a low-resource scenario where only 10 hours of data were used.
The models are then tested on the LibriSpeech \emph{test-clean} set.
In slot filling tasks, models are not only expected to recognize the spoken content but also to decode the slot type associated with that content.
Specifically, the slot type is decoded together with the transcription in a sequence generation manner~\cite{DBLP:conf/interspeech/YangCCLLLLSCLHT21, DBLP:conf/icassp/LaiCL0G21}.

In recent research~\cite{he2021towards}, prompting is also viewed as a parameter-efficient transfer learning method.
To evaluate the transferability and effectiveness of Wav2Seq with prompting, we conducted cross-lingual ASR experiments on 7 different languages: Dutch, French, German, Italian, Polish, Portuguese, and Spanish. For these experiments, we utilized the multilingual-LibriSpeech (MLS) dataset~\cite{DBLP:conf/interspeech/PratapXSSC20}. 
The Wav2Seq model was initially pre-trained on the English-only LibriSpeech-960 dataset~\cite{DBLP:conf/icassp/PanayotovCPK15}. 
By incorporating prompting, we aim to make the Wav2Seq transfer the knowledge to the target language and effectively perform cross-lingual ASR.
We explored different prompt lengths, specifically 60 and 120, and compared them to the fine-tuning method, where we fine-tuned the top 1 and 2 layers of the decoder.

\begin{table*}[t]
    
	\scriptsize\centering
	\caption{Prompting method on classification tasks and sequence generation tasks. \textbf{FT}: fine-tuning the pre-trained Wav2Seq. $\#$: The number of trainable parameters.} 
	\vskip -0.02in
	\resizebox{\linewidth}{!}{
	\begin{tabular}{ccccc@{\hspace{1.5em}}cccccc}
		\toprule
		\multirow{3}{*}{\textbf{Scenarios}}
		&\multicolumn{2}{c}{Intent Classification} & \multicolumn{2}{c}{Keyword Spotting} && ASR (100~hr) & ASR (10~hr) & \multicolumn{2}{c}{Slot Filling} \\
		\cmidrule{2-3}\cmidrule{4-5}\cmidrule{7-10} & Acc ($\uparrow$) &$\#$ & Acc ($\uparrow$) & $\#$ && WER ($\downarrow$) & WER ($\downarrow$) & CER ($\downarrow$) & F1 ($\uparrow$) & $\#$        \\
		\midrule
		Wav2Seq-Prompt         & \textbf{98.79}   & 0.2M  & \textbf{98.40}   & 0.2M && \textbf{9.28}              & \textbf{14.13}              & \textbf{9.69}      & \textbf{84.97}  & 4.3M        \\
		SpeechPrompt~\cite{chang2022exp}              & 98.40            & 0.15M & 95.16            & 0.08M && 34.17              & -                  & 59.47              & 66.90           & 4.5M \\
		\midrule
		Wav2Seq-FT         & 99.50            & 155M  & 98.20            & 155M && 5.57              & 10.20              & 4.65               & 93.21           & 155M \\
		\bottomrule
  \label{table:prompt_exp}
	\end{tabular}
        }
\end{table*}

\begin{table*}[h]
	\centering
\caption{Transfer learning results on cross-lingual ASR using prompting (Wav2Seq-Prompt) and conventional fine-tuning (Wav2Seq-FT) across 7 languages.
The fine-tuning of the top $n$ layers of the decoder and different prompt length ($L$) of prompting are presented.
We report word error rate (WER) and the number of trainable parameters for each scenario.}
	\vskip -0.05in
	\begin{minipage}[c!]{\linewidth}
		\centering
            \begin{tabular}{ccccccccc}
			\toprule
			\textbf{Scenario} & Dutch & French & German & Italian & Polish  & Portuguese & Spanish & $\#$ \\
			\midrule
                Wav2Seq-FT ($n = 1$) & 49.08 & 41.81 & 32.54 & 38.17 & 36.64 & 43.87 & 32.41 & 10.2M \\
                Wav2Seq-FT ($n = 2$) & 43.07 & 33.76 & 23.49 & 31.72 & \textbf{30.25} & 36.38 & 25.52 & 19.7M \\
                Wav2Seq-Prompt ($L = 60$) & 48.77	& 43.23	& 34.96	& 39.79	& 42.17	& 44.02	& 36.24 & 2.5M \\
                Wav2Seq-Prompt ($L = 120$) & \textbf{41.3}	& \textbf{32.48}	& \textbf{23.25}	& \textbf{29.28}	& 30.39	& \textbf{33.24} & \textbf{23.25} & 4.3M \\
			\bottomrule
			\end{tabular}
	\end{minipage}
    \label{table:adapter_exp}
\end{table*}


\begin{table}[h]
\begin{minipage}{\linewidth}
\caption{WER on German ASR when prompts are placed at some of the layers. E: encoder, D: decoder, f3: first 3 layers, $\ell$3: last 3 layers and so on. \#: trainable parameters.}
\scriptsize\centering
    \begin{tabular}{ccccccccc}
        \hline
        \multirow{2}{*}{\textbf{Placement}} & \multicolumn{4}{c}{Encoder} & \multicolumn{2}{c}{Decoder} & & \\
        \cmidrule(lr){2-5}\cmidrule(lr){6-7}
        {} & f3 & $\ell$3 & f6 & $\ell$6 & f3 & $\ell$3  & \# & WER 
 $\downarrow$\\
        \hline 
        E~(f3) & \checkmark & & & & & &  1.5M & 29.43 \\
        E~($\ell$3) & & \checkmark & & & & &  1.5M & \textbf{26.52}\\
        D~(f3) & & & & & \checkmark & &  1.5M & 28.32\\
        D ($\ell$3) & & & & & & \checkmark &  1.5M & 27.88\\
        \hline
        E~(f6) & & & \checkmark & & & &  2.1M & 28.08\\
        E~($\ell$6) & & & & \checkmark & & &  2.1M & \textbf{24.65}\\
        D~(all)  & & & & & & &   2.1M & 27.41\\
        E~(f3) + D~(f3) & \checkmark & & & & \checkmark & &  2.1M & 28.24\\
        E~(f3) + D~($\ell$3) & \checkmark & & & & & \checkmark &  2.1M & 28.98\\
        E~($\ell$3) + D~(f3) & & \checkmark & & & \checkmark & &  2.1M & 26.32\\
        E~($\ell$3) + D~($\ell$3) & & \checkmark & & & & \checkmark &  2.1M & 26.27\\    
        \hline
        E (all) + D(all) & & & \checkmark & \checkmark & \checkmark & \checkmark &  4.3M & 23.25\\    
        \hline
    \end{tabular}
\label{table:prompt_placement}
\end{minipage}
\end{table}
\subsection{Adapter tuning}
We further compared adapter tuning with prompting methods in the aforementioned cross-lingual speech recognition experiment.
The experiment was conducted in both the full-dataset scenario and the low-resource scenario.
In the low-resource scenario, only 10 hours of training data were provided for each language.
To ensure a fair comparison between prompting and adapter tuning, we aligned the trainable parameters of both methods.
In the low-resource scenario, we set the prompt length to 120, and the adapter bottleneck dimension to 64.
In the full-dataset setting, we increased the prompt length to 180 tokens, while the adapter bottleneck dimension was set to 96.

\begin{figure*}[t]
    \centering
    \includegraphics[width=\linewidth]{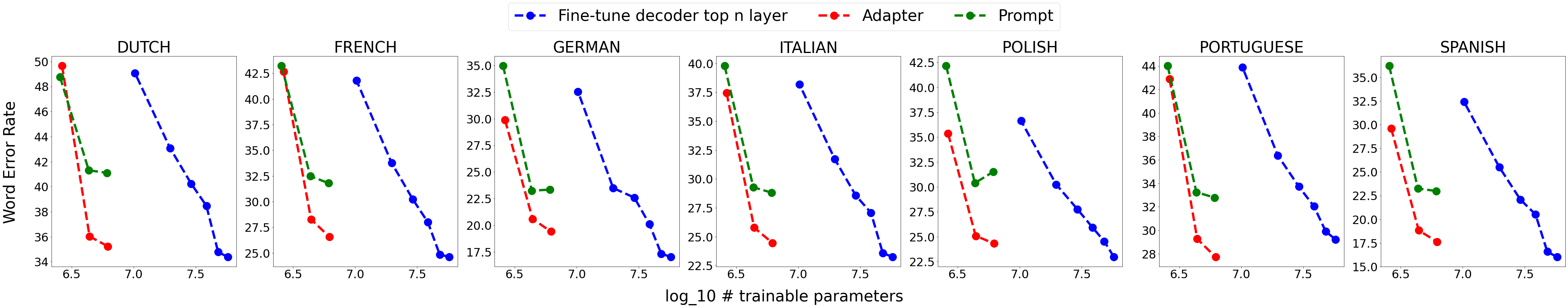}
    \caption{Comparison of prompting method, adapter tuning, and conventional fine-tuning methods.
    The red line represents the results with different adapter sizes (32, 64, 96).
    The green line represents the results of different prompt lengths (60, 120, 180).
    The blue line represents the results of training the top n decoder layers, where n ranges from 1 to 6.}
    \label{fig:all_methods}
\end{figure*}
\begin{table*}[h]
	\centering
        \caption{Comparing the performance of prompting and adapter tuning on multilingual ASR in both the full-dataset scenario and the low-resource scenario.}
	\vskip -0.05in
	\begin{minipage}[c!]{\linewidth}
		\centering
        \begin{tabular}{ccccccccccc}
			\toprule
			\multicolumn{2}{c}{\textbf{Scenarios}} & Dutch & French & German & Italian & Polish  & Portuguese & Spanish & $\#$ \\
			\midrule
			
			\multirow{2}{*}{\textbf{10h}} & 
                Wav2Seq-Prompt & \textbf{50.92} & \textbf{48.96} & \textbf{36.39} & \textbf{43.99} & \textbf{44.05} & \textbf{49.58} & \textbf{36.04} & 4.3M \\
                & Wav2Seq-Adapter & 52.90 & 66.84 & 38.13 & 58.78 & 57.47 & 68.14 & 46.04 & 4.4M \\
   
                \midrule\midrule
   
			\multirow{2}{*}{\textbf{full}} 
                & Wav2Seq-Prompt & 41.11 & 31.80 & 23.34 & 28.81 &	31.52 & 32.78 & 22.99 & 6.1M \\
               & Wav2Seq-Adapter & \textbf{35.23} & \textbf{26.56} & \textbf{19.41} & \textbf{24.46} & \textbf{24.32} & \textbf{27.72} & \textbf{17.62} & 6.2M \\
   
			\bottomrule
			\label{table:prompt_adapter_exp}
			\end{tabular}
	\end{minipage}
 \vspace{-0.1cm}
\end{table*}

\section{Results}
\subsection{Prompting for various speech processing tasks}
Table~\ref{table:prompt_exp} provides the results of prompting Wav2Seq, denoted as \textbf{Wav2Seq-Prompt}, and prompting GSLM in \textbf{SpeechPrompt} for both classification tasks and sequence generation tasks.
Our findings demonstrate that, within the prompting paradigm, Wav2Seq-Prompt consistently outperforms SpeechPrompt across all tasks. 
It's worth noting that in sequence generation tasks, Wav2Seq-Prompt exhibits significant improvements over SpeechPrompt.
For ASR, it achieves a remarkable 73\% reduction in WER, and for the slot filling (SF) task, it achieves an impressive 84\% reduction in CER and 27\% improvement in F1 score. This study showcases the successful application of the speech prompting paradigm for achieving impressive results in sequence generation tasks.

To assess the benefits of prompting in the low-resource scenario, we tested Wav2Seq on low-resource ASR. 
While SpeechPrompt failed to produce meaningful results, Wav2Seq demonstrated competitive performance.
Despite the limitations of low-resource data, Wav2Seq achieved impressive results, with the WER being only 4.85 higher compared to the high-resource scenario. 
We believe that the outstanding performance of Wav2Seq-Prompt in sequence generation tasks is attributed to its pseudo speech recognition pre-training task and its encoder-decoder architecture.
On the other hand, the GSLM in SpeechPrompt is pre-trained with speech continuation task, which differs from the sequence-to-sequence nature of ASR and slot filling.

We also present the performance results of fine-tuning the entire Wav2Seq model (referred to as \textbf{Wav2Seq-FT}) as a performance upper bound for comparison.
Encouragingly, our results indicate that prompting on Wav2Seq does not suffer from significant performance degradation compared to Wav2Seq-FT.
In fact, Wav2Seq-Prompt outperforms Wav2Seq-FT in the KS task. 
These remarkable achievements are accomplished by utilizing within 3\% of the trainable parameters out of the entire model.

\subsection{Prompting for cross-lingual transfer learning}
Table~\ref{table:adapter_exp} shows the results for cross-lingual ASR across 7 languages.
The pre-trained Wav2Seq model, initially trained on an English-only dataset, is tested to assess the transferability of the prompting method.
The results show that as the number of trainable parameters grows, both FT and prompting exhibit improved performance (reduced WER).
However, prompting achieves better parameter efficiency.
When the prompt length is 120, Wav2Seq-Prompt outperforms Wav2Seq-FT ($n = 2$) on 6 out of 7 languages, while using only about 20\% (4.3M out of 19.7M) of the trainable parameters.

Prompts can be integrated at different input positions within the Transformer layers.
We further explore the placement of prompts in the pre-trained encoder-decoder model. 
As a case study, we employ cross-lingual ASR on German and explore various prompt positions within both the encoder and decoder layers.
Table.~\ref{table:prompt_placement} presents a comprehensive analysis.
We find that incorporating the prompts at the last layers of the encoder yields the best effectiveness.
For example, adding prompts at the last 3 encoder layers (\textbf{E($\bm{\ell}$3)}) or the last 6 encoder layers (\textbf{E($\bm{\ell}$6)}) achieves the lowest WER compared to other methods with the same number of trainable parameters.
Further investigation into the underlying reasons behind these observations is left for future work.

\subsection{Prompting and adapter tuning as parameter-efficient transfer learning methods}
Recent study~\cite{he2021towards} has also considered the prompt tuning method as an adapter, recognizing it as a parameter-efficient transfer learning technique. 
To comprehensively understand the transferability of prompting, adapter tuning, and fine-tuning methods, we present the cross-lingual ASR results in Fig.~\ref{fig:all_methods}.
We found that prompting and adapter tuning can achieve better parameter efficiency compared to fine-tuning.
When the performance levels are similar, both prompting and adapter tuning require fewer trainable parameters.
We also found that when the number of trainable parameters increases, the prompting method does not always increase performance.
In the German and Polish ASR tasks, we observed that using a prompt length of 180 resulted in underperformance compared to using a prompt length of 120.
We believe adapter tuning offers more model capacity when the trainable parameters are large while prompting has its limit.

To further investigate prompting and adapter tuning, we align the trainable parameters of them and evaluate their performance in cross-lingual ASR tasks in low-resource and full-dataset scenarios.
Table~\ref{table:prompt_adapter_exp} shows the results.
Surprisingly, we observe different trends in prompting and adapter tuning.
In the low-resource scenario, the prompting method consistently outperforms the adapter-based method across all languages.
However, in the full-dataset scenario, the adapter method surpasses the performance of the prompting method.

Combining the results from Table~\ref{table:prompt_adapter_exp} and Fig. ~\ref{fig:all_methods}, we conclude that both prompting and adapter tuning outperform fine-tuning method.
Prompting outperforms adapter tuning in the low-resource scenario.
This is probably because prompting does not modify the model architecture, resulting in less disruption to existing knowledge of the model.
However, when the training data and trainable parameter budget are enough, adapter tuning becomes more effective for transfer learning.

\section{Conclusions}
\label{sec:conclusions}
The increase in model sizes results in a growing demand for computational resource-friendly approaches to task adaptation.
This paper presents the first application of prompting and adapter tuning techniques to the SSL encoder-decoder speech model.
Encouragingly, the experimental results show the proposed method delivers substantial improvements in sequence generation tasks compared to SpeechPrompt.
Also, these approaches show remarkable parameter efficiency when compared to fine-tuning methods in the cross-lingual transfer learning scenario.
We also discovered that in the low-resource scenario, prompting outperforms adapter tuning.

\section{Acknowledgement}
We thank the National Center for High-performance Computing (NCHC) of National Applied Research Laboratories (NARLabs) in Taiwan for providing computational and storage resources.

\bibliographystyle{IEEEbib}
\bibliography{refs}

\begin{thebibliography}{10}

\bibitem{oord2018representation}
Aaron van~den Oord, Yazhe Li, and Oriol Vinyals,
\newblock ``Representation learning with contrastive predictive coding,''
\newblock {\em arXiv preprint arXiv:1807.03748}, 2018.

\bibitem{baevski2020wav2vec}
Alexei Baevski, Yuhao Zhou, Abdelrahman Mohamed, and Michael Auli,
\newblock ``wav2vec 2.0: A framework for self-supervised learning of speech
  representations,''
\newblock {\em Advances in neural information processing systems}, vol. 33, pp.
  12449--12460, 2020.

\bibitem{hsu2021hubert}
Wei-Ning Hsu, Benjamin Bolte, Yao-Hung~Hubert Tsai, Kushal Lakhotia, Ruslan
  Salakhutdinov, and Abdelrahman Mohamed,
\newblock ``Hubert: Self-supervised speech representation learning by masked
  prediction of hidden units,''
\newblock {\em IEEE/ACM Transactions on Audio, Speech, and Language
  Processing}, vol. 29, pp. 3451--3460, 2021.

\bibitem{DBLP:conf/interspeech/YangCCLLLLSCLHT21}
Shu{-}Wen Yang et~al.,
\newblock ``{SUPERB:} speech processing universal performance benchmark,''
\newblock in {\em Interspeech}. 2021, pp. 1194--1198, {ISCA}.

\bibitem{mohamed2022self}
Abdelrahman Mohamed et~al.,
\newblock ``Self-supervised speech representation learning: A review,''
\newblock {\em IEEE Journal of Selected Topics in Signal Processing}, 2022.

\bibitem{lakhotia2021generative}
Kushal Lakhotia et~al.,
\newblock ``On generative spoken language modeling from raw audio,''
\newblock {\em Transactions of the Association for Computational Linguistics},
  vol. 9, pp. 1336--1354, 2021.

\bibitem{kharitonov2022text}
Eugene Kharitonov et~al.,
\newblock ``Text-free prosody-aware generative spoken language modeling,''
\newblock in {\em Proceedings of the 60th Annual Meeting of the Association for
  Computational Linguistics (Volume 1: Long Papers)}, 2022, pp. 8666--8681.

\bibitem{hassid2023textually}
Michael Hassid et~al.,
\newblock ``Textually pretrained speech language models,''
\newblock {\em arXiv preprint arXiv:2305.13009}, 2023.

\bibitem{wu2023wav2seq}
Felix Wu et~al.,
\newblock ``Wav2seq: Pre-training speech-to-text encoder-decoder models using
  pseudo languages,''
\newblock in {\em ICASSP 2023-2023 IEEE International Conference on Acoustics,
  Speech and Signal Processing (ICASSP)}. IEEE, 2023, pp. 1--5.

\bibitem{liu2023pre}
Pengfei Liu, Weizhe Yuan, Jinlan Fu, Zhengbao Jiang, Hiroaki Hayashi, and
  Graham Neubig,
\newblock ``Pre-train, prompt, and predict: A systematic survey of prompting
  methods in natural language processing,''
\newblock {\em ACM Computing Surveys}, vol. 55, no. 9, pp. 1--35, 2023.

\bibitem{houlsby2019parameter}
Neil Houlsby, Andrei Giurgiu, Stanislaw Jastrzebski, Bruna Morrone, Quentin
  De~Laroussilhe, Andrea Gesmundo, Mona Attariyan, and Sylvain Gelly,
\newblock ``Parameter-efficient transfer learning for nlp,''
\newblock in {\em International Conference on Machine Learning}, 2019, pp.
  2790--2799.

\bibitem{chang2022exp}
Kai-Wei Chang, Wei-Cheng Tseng, Shang-Wen Li, and Hung yi~Lee,
\newblock ``{An Exploration of Prompt Tuning on Generative Spoken Language
  Model for Speech Processing Tasks},''
\newblock in {\em Proc. Interspeech 2022}, 2022, pp. 5005--5009.

\bibitem{chang2023speechprompt}
Kai-Wei Chang et~al.,
\newblock ``{SpeechPrompt} v2: Prompt tuning for speech classification tasks,''
\newblock {\em arXiv preprint arXiv:2303.00733}, 2023.

\bibitem{otake2023parameter}
Shinta Otake, Rei Kawakami, and Nakamasa Inoue,
\newblock ``Parameter efficient transfer learning for various speech processing
  tasks,''
\newblock in {\em ICASSP 2023-2023 IEEE International Conference on Acoustics,
  Speech and Signal Processing (ICASSP)}. IEEE, 2023, pp. 1--5.

\bibitem{chen2023exploring}
Zih-Ching Chen, Chin-Lun Fu, Chih-Ying Liu, Shang-Wen~Daniel Li, and Hung-yi
  Lee,
\newblock ``Exploring efficient-tuning methods in self-supervised speech
  models,''
\newblock in {\em 2022 IEEE Spoken Language Technology Workshop (SLT)}. IEEE,
  2023, pp. 1120--1127.

\bibitem{chen2022wavlm}
Sanyuan Chen et~al.,
\newblock ``Wavlm: Large-scale self-supervised pre-training for full stack
  speech processing,''
\newblock {\em IEEE Journal of Selected Topics in Signal Processing}, vol. 16,
  no. 6, pp. 1505--1518, 2022.

\bibitem{ao2022pre}
Junyi Ao et~al.,
\newblock ``Pre-training transformer decoder for end-to-end asr model with
  unpaired speech data,''
\newblock {\em arXiv preprint arXiv:2203.17113}, 2022.

\bibitem{borsos2023audiolm}
Zal{\'a}n Borsos et~al.,
\newblock ``Audiolm: a language modeling approach to audio generation,''
\newblock {\em IEEE/ACM Transactions on Audio, Speech, and Language
  Processing}, 2023.

\bibitem{DBLP:conf/interspeech/ArunkumarU22}
A.~Arunkumar and Srinivasan Umesh,
\newblock ``Joint encoder-decoder self-supervised pre-training for {ASR},''
\newblock in {\em {INTERSPEECH}}. 2022, pp. 3418--3422, {ISCA}.

\bibitem{petroni2019language}
Fabio Petroni, Tim Rockt{\"a}schel, Sebastian Riedel, Patrick Lewis, Anton
  Bakhtin, Yuxiang Wu, and Alexander Miller,
\newblock ``Language models as knowledge bases?,''
\newblock in {\em Proceedings of the 2019 Conference on Empirical Methods in
  Natural Language Processing and the 9th International Joint Conference on
  Natural Language Processing (EMNLP-IJCNLP)}, 2019, pp. 2463--2473.

\bibitem{shin2020autoprompt}
Taylor Shin, Yasaman Razeghi, Robert~L Logan~IV, Eric Wallace, and Sameer
  Singh,
\newblock ``Autoprompt: Eliciting knowledge from language models with
  automatically generated prompts,''
\newblock in {\em Proceedings of the 2020 Conference on Empirical Methods in
  Natural Language Processing (EMNLP)}, 2020, pp. 4222--4235.

\bibitem{li2021prefix}
Xiang~Lisa Li and Percy Liang,
\newblock ``Prefix-tuning: Optimizing continuous prompts for generation,''
\newblock in {\em Proceedings of the 59th Annual Meeting of the Association for
  Computational Linguistics and the 11th International Joint Conference on
  Natural Language Processing (Volume 1: Long Papers)}, 2021, pp. 4582--4597.

\bibitem{liu2022p}
Xiao Liu, Kaixuan Ji, Yicheng Fu, Weng Tam, Zhengxiao Du, Zhilin Yang, and Jie
  Tang,
\newblock ``P-tuning: Prompt tuning can be comparable to fine-tuning across
  scales and tasks,''
\newblock in {\em Proceedings of the 60th Annual Meeting of the Association for
  Computational Linguistics (Volume 2: Short Papers)}, 2022, pp. 61--68.

\bibitem{wu2023speechgen}
Haibin Wu, Kai-Wei Chang, Yuan-Kuei Wu, and Hung-yi Lee,
\newblock ``{SpeechGen}: Unlocking the generative power of speech language
  models with prompts,''
\newblock {\em arXiv preprint arXiv:2306.02207}, 2023.

\bibitem{DBLP:conf/iclr/ElsayedGS19}
Gamaleldin~F. Elsayed, Ian~J. Goodfellow, and Jascha Sohl{-}Dickstein,
\newblock ``Adversarial reprogramming of neural networks,''
\newblock in {\em {ICLR} (Poster)}. 2019, OpenReview.net.

\bibitem{DBLP:journals/corr/abs-2202-10629}
Pin{-}Yu Chen,
\newblock ``Model reprogramming: Resource-efficient cross-domain machine
  learning,''
\newblock {\em CoRR}, vol. abs/2202.10629, 2022.

\bibitem{yen2021neural}
Hao Yen, Pin-Jui Ku, Chao-Han~Huck Yang, Hu~Hu, Sabato~Marco Siniscalchi,
  Pin-Yu Chen, and Yu~Tsao,
\newblock ``Neural model reprogramming with similarity based mapping for
  low-resource spoken command classification,''
\newblock {\em arXiv preprint arXiv:2110.03894}, 2021.

\bibitem{yang2023english}
Chao-Han~Huck Yang, Bo~Li, Yu~Zhang, Nanxin Chen, Rohit Prabhavalkar, Tara~N
  Sainath, and Trevor Strohman,
\newblock ``From english to more languages: Parameter-efficient model
  reprogramming for cross-lingual speech recognition,''
\newblock {\em arXiv preprint arXiv:2301.07851}, 2023.

\bibitem{hung2023low}
Yun-Ning Hung, Chao-Han~Huck Yang, Pin-Yu Chen, and Alexander Lerch,
\newblock ``Low-resource music genre classification with cross-modal neural
  model reprogramming,''
\newblock in {\em ICASSP 2023-2023 IEEE International Conference on Acoustics,
  Speech and Signal Processing (ICASSP)}. IEEE, 2023, pp. 1--5.

\bibitem{rebuffi2017learning}
Sylvestre-Alvise Rebuffi, Hakan Bilen, and Andrea Vedaldi,
\newblock ``Learning multiple visual domains with residual adapters,''
\newblock {\em Advances in neural information processing systems}, vol. 30,
  2017.

\bibitem{DBLP:conf/iclr/HuSWALWWC22}
Edward~J. Hu, Yelong Shen, Phillip Wallis, Zeyuan Allen{-}Zhu, Yuanzhi Li,
  Shean Wang, Lu~Wang, and Weizhu Chen,
\newblock ``Lora: Low-rank adaptation of large language models,''
\newblock in {\em {ICLR}}. 2022, OpenReview.net.

\bibitem{fu2022adapterbias}
Chin-Lun Fu, Zih-Ching Chen, Yun-Ru Lee, and Hung-Yi Lee,
\newblock ``Adapterbias: Parameter-efficient token-dependent representation
  shift for adapters in nlp tasks,''
\newblock in {\em Findings of the Association for Computational Linguistics:
  NAACL 2022}, 2022, pp. 2608--2621.

\bibitem{le2021lightweight}
Hang Le, Juan Pino, Changhan Wang, Jiatao Gu, Didier Schwab, and Laurent
  Besacier,
\newblock ``Lightweight adapter tuning for multilingual speech translation,''
\newblock in {\em Proceedings of the 59th Annual Meeting of the Association for
  Computational Linguistics and the 11th International Joint Conference on
  Natural Language Processing (Volume 2: Short Papers)}, 2021, pp. 817--824.

\bibitem{morioka2022residual}
Nobuyuki Morioka, Heiga Zen, Nanxin Chen, Yu~Zhang, and Yifan Ding,
\newblock ``Residual adapters for few-shot text-to-speech speaker adaptation,''
\newblock {\em arXiv preprint arXiv:2210.15868}, 2022.

\bibitem{vaswani2017attention}
Ashish Vaswani et~al.,
\newblock ``Attention is all you need,''
\newblock {\em Advances in neural information processing systems}, vol. 30,
  2017.

\bibitem{DBLP:journals/corr/abs-1804-03209}
Pete Warden,
\newblock ``Speech commands: {A} dataset for limited-vocabulary speech
  recognition,''
\newblock {\em CoRR}, vol. abs/1804.03209, 2018.

\bibitem{DBLP:conf/interspeech/LugoschRITB19}
Loren Lugosch, Mirco Ravanelli, Patrick Ignoto, Vikrant~Singh Tomar, and Yoshua
  Bengio,
\newblock ``Speech model pre-training for end-to-end spoken language
  understanding,''
\newblock in {\em {INTERSPEECH}}. 2019, pp. 814--818, {ISCA}.

\bibitem{DBLP:conf/icassp/PanayotovCPK15}
Vassil Panayotov, Guoguo Chen, Daniel Povey, and Sanjeev Khudanpur,
\newblock ``Librispeech: An {ASR} corpus based on public domain audio books,''
\newblock in {\em {ICASSP}}. 2015, pp. 5206--5210, {IEEE}.

\bibitem{librilight}
J.~{Kahn} et~al.,
\newblock ``Libri-light: A benchmark for asr with limited or no supervision,''
\newblock in {\em ICASSP 2020 - 2020 IEEE International Conference on
  Acoustics, Speech and Signal Processing (ICASSP)}, 2020, pp. 7669--7673,
\newblock \url{https://github.com/facebookresearch/libri-light}.

\bibitem{DBLP:conf/icassp/LaiCL0G21}
Cheng{-}I Lai, Yung{-}Sung Chuang, Hung{-}Yi Lee, Shang{-}Wen Li, and James~R.
  Glass,
\newblock ``Semi-supervised spoken language understanding via self-supervised
  speech and language model pretraining,''
\newblock in {\em {ICASSP}}. 2021, pp. 7468--7472, {IEEE}.

\bibitem{DBLP:conf/interspeech/PratapXSSC20}
Vineel Pratap, Qiantong Xu, Anuroop Sriram, Gabriel Synnaeve, and Ronan
  Collobert,
\newblock ``{MLS:} {A} large-scale multilingual dataset for speech research,''
\newblock in {\em {INTERSPEECH}}. 2020, pp. 2757--2761, {ISCA}.

\bibitem{he2021towards}
Junxian He, Chunting Zhou, Xuezhe Ma, Taylor Berg-Kirkpatrick, and Graham
  Neubig,
\newblock ``Towards a unified view of parameter-efficient transfer learning,''
\newblock in {\em International Conference on Learning Representations}, 2021.

\end{thebibliography}

\end{document}